\documentclass[pss]{wiley2sp} % provides pss two-column style
\usepackage{amsmath}
\usepackage{bm}        
\usepackage{comment}
\usepackage{color}

\begin{document}

% Title of the article
\title{Electronic States of Al-Mg-Zn Quasicrystal and Its Approximant based on the First-Principles Calculations}

% Authors
\author{%
  Masaki Saito,
  Takuya Sekikawa, and 
  Yoshiaki \={O}no\textsuperscript{\Ast}
}

%E-mail-address of corresponding author
\mail{e-mail
  \textsf{y.ono@phys.sc.niigata-u.ac.jp}}

% author's affiliations/addresses
\institute{%
Department of Physics, Niigata University, Niigata, 950-2181, Japan
}

% Please select about four verbal keywords for your manuscript.
\keywords{Quasicrystal, Approximant, First-principles calculation, Electronic state, Superconductivity}

\abstract{\bf%

First-principles calculations are performed to investigate the electronic states of 1/1 and 2/1 approximants with the composition Al$_{15}$Zn$_{40}$Mg$_{45}$ which is close to the quasicrystal (QC) Al$_{14.9}$Zn$_{41.0}$Mg$_{44.1}$ in which the superconductivity is recently discovered. The density of states for the 1/1 approximant shows a wide pseudogap structure near the Fermi level as commonly observed in various approximants of the QCs, whereas those for the 2/1 approximant do not show such a wide pseudogap. Instead of the wide pseudogap, the 2/1 approximant shows a remarkable narrow pseudogap at the Fermi level in contrast to the 1/1 approximant, which shows a shallow hump at the Fermi level within the wide pseudogap. This seems to be consistent with the experimental observation of the electrical resistivity, which increases with decreasing temperature from the room temperature down to around the superconducting transition temperature for the 2/1 approximant together with the QC, whereas it monotonically decreases for the 1/1 approximant.
}

\maketitle   % please do not remove

\section{Introduction}
In the classical crystallography, solids are conventionally divided into two classes: crystal and amorphous. The crystal has translational symmetry with twofold (or three-, four-, and sixfold) rotational symmetry, whereas the amorphous has no such symmetry. In contrast to the classical crystallography, quasicrystals (QCs), first discovered by Shechtman {\it et al.} in 1984 in Al-Mn alloy\cite{Shechtman}, have specific structures, which are not periodic but have kinds of translational orders with the fivefold (or eight-, ten-, and twelvefold) rotational symmetry and, then, are classified as the third class of solids. Because of the specific structures, QCs are expected to show unique physical properties and have been extensively investigated\cite{Stadnik}. 

Recently, superconductivity in the Al$_{14.9}$Zn$_{41.0}$Mg$_{44.1}$ QC at a transition temperature $T_c \sim 0.05$K has been discovered by Kamiya {\it et al.}\cite{Kamiya} and has attracted much attention for its symmetry and pairing mechanism. They have also found the superconductivity in its approximant crystals (ACs), which are periodic in contrast to QCs but have the local structures similar to the QC, with the two types of 1/1 and 2/1 ACs\cite{Kamiya}. In the 2/1 AC whose local structure is very close to the QC, the temperature dependence of the electrical resistivity $\rho$ almost coincides with that in the QC, where $\rho$ increases with decreasing temperature from the room temperature down to around $T_c$. On the other hand, in the 1/1 AC whose local structure is less similar to the QC than the 2/1 AC, $\rho$ monotonically decreases with decreasing temperature. Therefore, it is important to clarify the difference in the electronic states between the 1/1 and 2/1 ACs with the composition corresponding to the superconducting QC to discuss transport properties not only for the ACs but also for the QC. 

Theoretically, Sakai {\it et al.}\cite{Sakai} have recently investigated the superconductivity in the attractive Hubbard model on the two-dimensional Penrose lattice simulating a QC using the real-space dynamical mean-field theory and have found that spatially extended Cooper pairs appear due to the aperiodicity of the lattice. As for the electronic states, the 1/1 AC with the composition Al$_{30}$Zn$_{30}$Mg$_{40}$ was previously studied on the basis of the first-principles calculation, and its density of states (DOS) was found to show a wide pseudogap structure at the Fermi level as commonly observed in various ACs\cite{Sato}. However, the electronic states with the composition corresponding to the superconducting QC in addition to the 2/1 AC were not discussed there. 
The purpose of this article is to investigate the electronic states of the both 1/1 and 2/1 ACs with the composition Al$_{15}$Zn$_{40}$Mg$_{45}$ close to the superconducting Al$_{14.9}$Zn$_{41.0}$Mg$_{44.1}$ QC and to clarify the difference between the 1/1 and 2/1 ACs.

\section{Calculation Method}

We perform the first-principles calculations for the 1/1 and 2/1 ACs with  the composition  Al$_{15}$Zn$_{40}$Mg$_{45}$ using OpenMX code\cite{OpenMX}, which is a fully linear-scaling density functional theory (DFT) method both in the building of the DFT Hamiltonian and in its solution. We use norm-conserved pseudopotentials\cite{Troullier} with an energy cutoff of 200Ry for charge density and pseudo-atomic localized functions\cite{Ozaki}, which are set as Al7.0-s3p3d2, Zn6.0-s3p3d3f1 and Mg7.0-s3p3d2\cite{Zn6.0}. We also set k-space sampling points of 7 $\times$ 7 $\times$ 7 for reciprocal lattice vectors.

\begin{figure}[t]%
\includegraphics[width=.5\textwidth, bb=0 0 241 241]{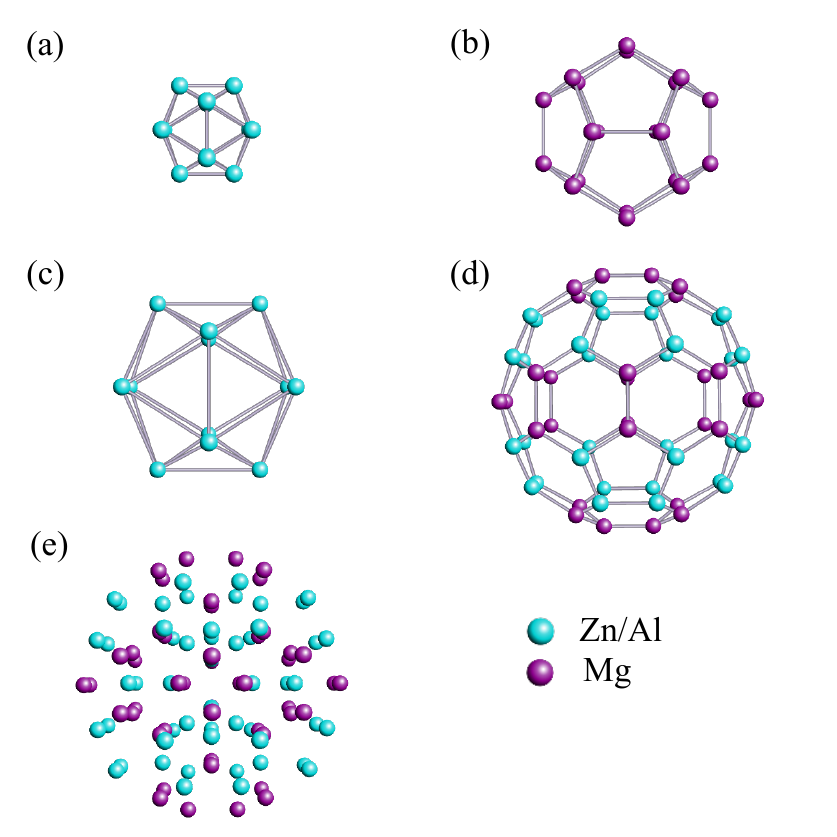}
\caption{%
 Bergman-type cluster in the Al-Zn-Mg systems consists of four shells with the first and third icosahedral shells composed by 12 Zn/Al atoms (a) and (c), the second dodecahedral shell composed of 20 Mg atoms (b) and the fourth truncated icosahedral shell composed of 60 atoms of Mg and Zn/Al (d), together with  the four stacked shells composed of the all atoms (e).  
}
\label{fig1}
\end{figure}

The basic structure in the Al-Zn-Mg QC and AC systems is the Bergman-type cluster\cite{Bergman} consists of four shells as shown in Figures \ref{fig1} (a)-(e). The first and third shells are icosahedrons composed of 12 Zn/Al atoms, the second shell is a dodecahedron composed of 20 Mg atoms, and the fourth shell is a truncated icosahedron composed of 60 atoms of Mg and Zn/Al, where chemical disorder in Zn/Al atoms possibly occurs. We note that the atoms of the fourth shell of a cluster are partially shared with the fourth shells of adjacent clusters, and then, the number of atoms of the fourth shell in a primitive unit cell is 36. The X-ray diffraction measurements\cite{Kamiya} indicate that the space group for the 1/1 Al$_{14.9}$Zn$_{41.0}$Mg$_{44.1}$ AC is $Im\bar{3}$ with the lattice parameter a$_{1/1}=14.195$\AA \ and that for the 2/1 Al$_{14.9}$Zn$_{42.1}$Mg$_{43.0}$ AC is $Pa\bar{3}$ with a$_{2/1}=23.006$\AA. 

As it is difficult to consider the chemical disorder within the first-principles calculations, we assume an ordered configuration of Zn/Al atoms, where the first (third) shell is exclusively composed of 12 Zn (Al) atoms, the second shell is composed of 20 Mg atoms, and the forth shell is composed of 28 Mg atoms and 32 Zn atoms, as shown in Figures \ref{fig1} (a)-(d). Among the 60 atoms of the fourth shell, 36 atoms with 16 Mg atoms and 20 Zn atoms are included in a primitive unit cell as noted before, and then, a primitive unit cell consists of 12 Al atoms, 32 Zn atoms and 36 Mg atoms in total corresponding to the composition Al$_{15}$Zn$_{40}$Mg$_{45}$. We also use the experimentally determined space groups and lattice parameters of the 1/1 Al$_{14.9}$Zn$_{41.0}$Mg$_{44.1}$ AC and the 2/1 Al$_{14.9}$Zn$_{42.1}$Mg$_{43.0}$ AC mentioned before for calculations of the 1/1 and 2/1 Al$_{15}$Zn$_{40}$Mg$_{45}$ ACs, respectively.

\section{Result}
\begin{figure*}[t]%
\subfloat{%
\includegraphics*[width=8.5cm, height=12cm, bb=0 0 241 340]{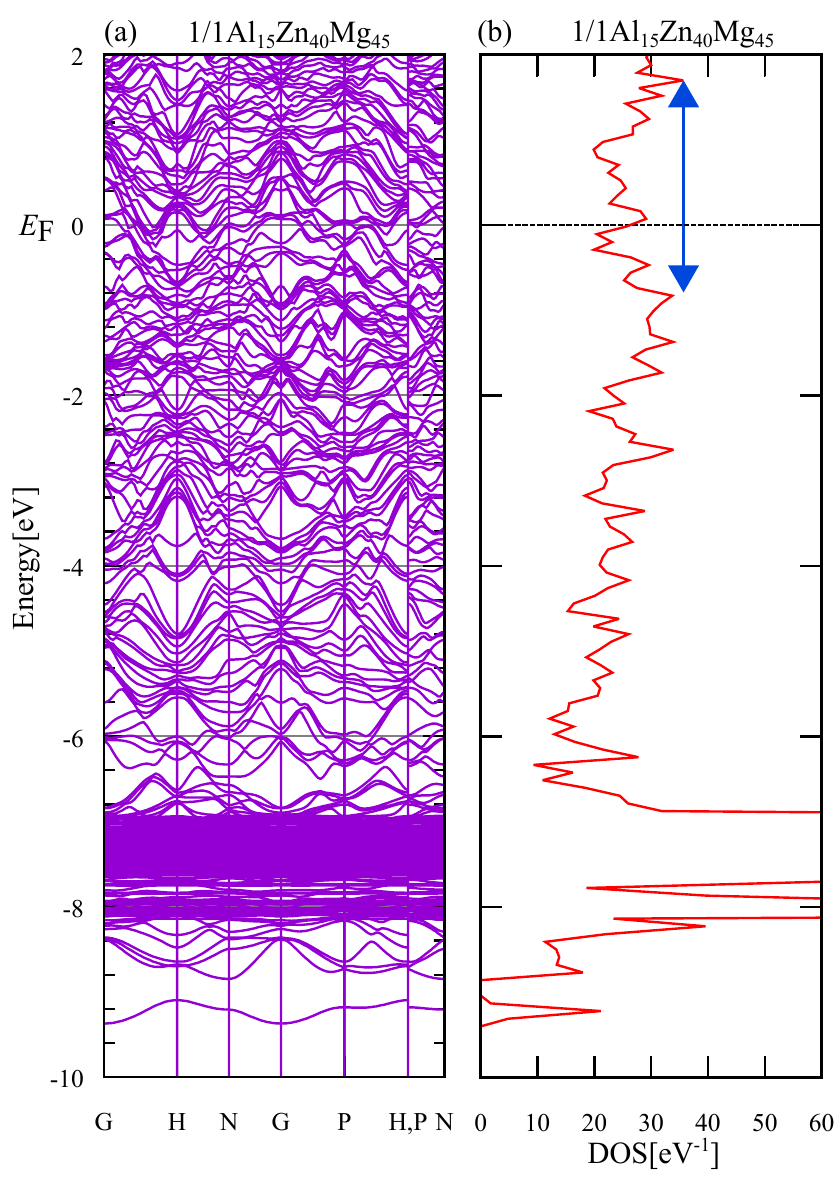}}\hfill
\subfloat{%
\includegraphics*[width=8.5cm, height=12cm, bb=0 0 241 340]{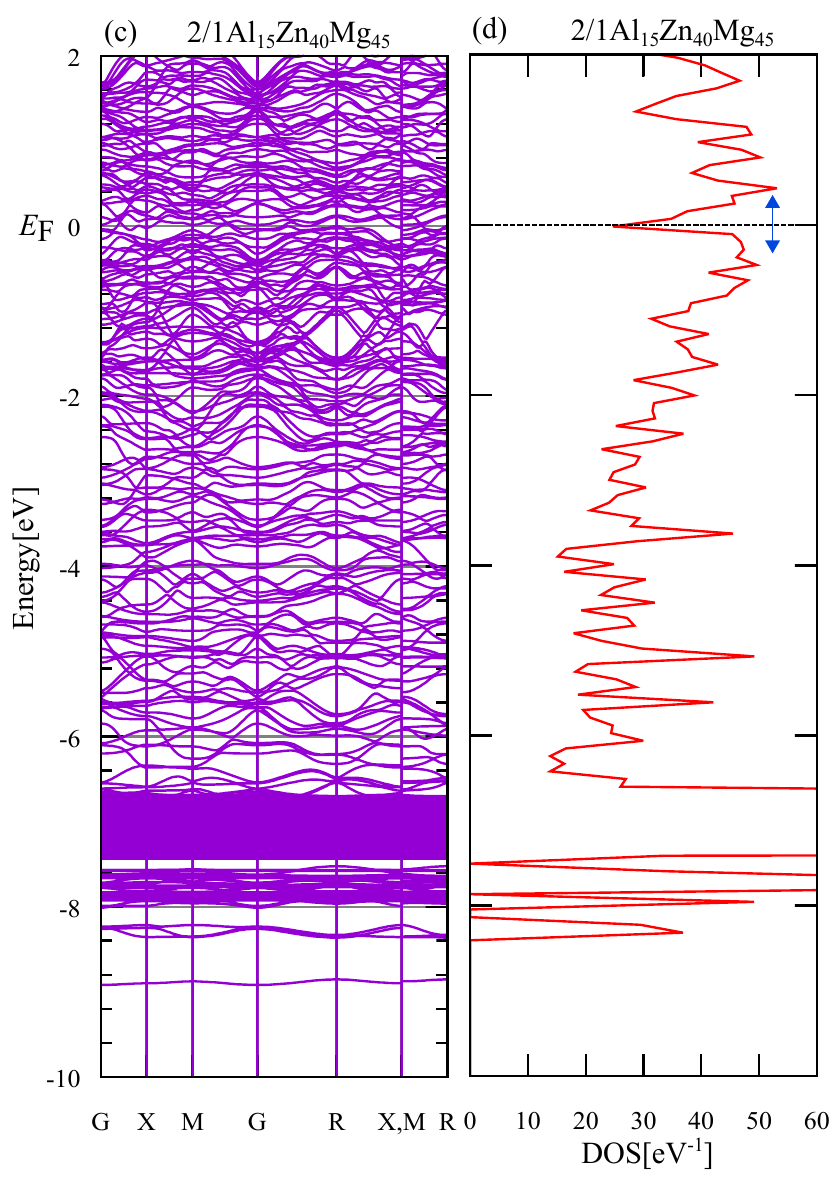}}\hfill
\caption{%
The energy band dispersions and the density of states for the 1/1 AC (a) and (b) and those for the 2/1 AC (c) and (d) with the composition Al$_{15}$Zn$_{40}$Mg$_{45}$ obtained from the first-principles calculations (OpenMX),  where $E_{\rm F}=0$ is the Fermi level and the two-way arrows show the wide and the narrow pseudogaps for the 1/1 AC in (b) and for the 2/1 AC in (d), respectively.   
}
\label{fig2}
\end{figure*}

Figures \ref{fig2} (a)-(d) show the energy band dispersions and the densities of states for the 1/1 and 2/1 ACs with the composition  Al$_{15}$Zn$_{40}$Mg$_{45}$. In the case with the 1/1 AC, we observe a wide pseudogap structure with the width of $O$(1eV) in the DOS near the Fermi level (see Figure \ref{fig2} (b)) as previously observed in the 1/1 AC with the different composition Al$_{30}$Zn$_{30}$Mg$_{40}$\cite{Sato}. As is well known, the pseudogap structures have been commonly observed in various ACs and have frequently discussed in relation to the stability of QCs\cite{Stadnik}. On the other hand, the 2/1 AC does not show such a wide pseudo gap but shows a remarkable narrow pseudogap with the width of $O$(0.1eV) at the Fermi level (see Figure \ref{fig2} (d)). This is a striking contrast to the 1/1 AC, which shows a shallow hump with the width of $O$(0.1eV) at the Fermi level within the wide pseudogap (see Figure \ref{fig2} (b)). These results seem to be consistent with the experimental observation of the electrical resistivity, which increases with decreasing temperature from the room temperature down to around $T_c$ for the 2/1 AC, whereas it monotonically decreases for the 1/1 AC as mentioned before\cite{Kamiya}.

\begin{figure}[t]%
\includegraphics[width=8cm, height=8cm, bb=0 0 198 227]{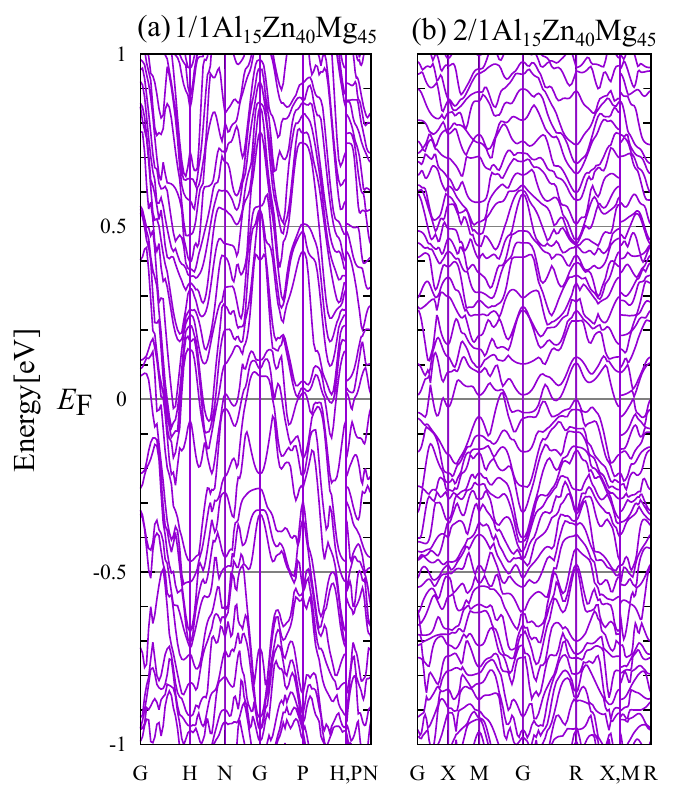}
\caption{%
The energy band dispersions near the Fermi level in the narrow range of -1 to 1 eV for the 1/1 AC (a) and those for the 2/1 AC (b).
}
\label{fig3}
\end{figure}

Due to the complex structure of the Bergman-type cluster composed of many atoms (see Figure \ref{fig1}), the energy bands are complexly intertwined, and then, many energy bands cross the Fermi level, as shown in Figures \ref{fig3} (a) and (b). In the case with the 1/1 AC, the Fermi level sits in around the middles of the relatively wide energy bands across the Fermi level, resulting in the shallow hump in the DOS, as shown in  Figure \ref{fig2} (b) and Figure \ref{fig3} (a). On the other hand, in the case with the 2/1 AC, the Fermi level sits in around the edges of the relatively narrow energy bands across the Fermi level, resulting in a semiconducting behavior with the narrow pseudogap together with the relatively large DOS around the Fermi level, as shown in Figure \ref{fig2} (d) and Figure \ref{fig3} (b).

We also estimate the specific heat coefficient $\gamma$ from the DOS per spin at the Fermi level $D(E_{\rm F})$ using the relation 
\begin{equation*}
\gamma = \frac{2\pi^2 k_{\rm B}^2}{3}D(E_{\rm F}). 
\end{equation*}
The obtained values of $\gamma$ are 0.778mJ K$^{-2}$mol$^{-1}$ for the 1/1 AC and 0.793mJ K$^{-2}$mol$^{-1}$ for the 2/1 AC. Those are comparable to the experimental values of $\gamma$, which are 0.813mJ K$^{-2}$mol$^{-1}$ for the 1/1 Al$_{14.9}$Zn$_{41.0}$Mg$_{44.1}$ AC and 0.726mJ K$^{-2}$mol$^{-1}$ for the 2/1 Al$_{14.9}$Zn$_{42.1}$Mg$_{43.0}$ AC\cite{Kamiya}, although the compositions are slightly different from Al$_{15}$Zn$_{40}$Mg$_{45}$ for the both 1/1 and 2/1 ACs in the present calculations. The calculated value of $\gamma$ for the 1/1 AC is slightly smaller than that for the 2/1 AC, in contrast to the experimental observation where $\gamma$ for the 1/1 AC is larger than that for the 2/1 AC. This discrepancy will be discussed later.

\section{Summary and discussions}

In summary, we have investigated the electronic states of 1/1 and 2/1 ACs with the composition Al$_{15}$Zn$_{40}$Mg$_{45}$, which is close to the superconducting QC Al$_{14.9}$Zn$_{41.0}$Mg$_{44.1}$ on the basis of the first-principles calculation. What we have found are:   1) The DOS for the 1/1 AC shows a wide pseudogap structure with the width of $O$(1eV) near the Fermi level as commonly observed in various ACs, whereas those for the 2/1 AC do not show such a wide pseudogap. 2) Instead of the wide pseudogap, the 2/1 AC shows a remarkable narrow pseudogap with the width of $O$(0.1eV) at the Fermi level in contrast to the 1/1 AC, which shows a shallow hump at the Fermi level within the wide pseudogap. 3) The calculated values of $\gamma$, 0.778mJ K$^{-2}$mol$^{-1}$ for the 1/1 AC and 0.793mJ K$^{-2}$mol$^{-1}$ for the 2/1 AC, are comparable to the experimental values of $\gamma$, 0.813mJ K$^{-2}$mol$^{-1}$ for the 1/1 Al$_{14.9}$Zn$_{41.0}$Mg$_{44.1}$ AC and 0.726mJ K$^{-2}$mol$^{-1}$ for the 2/1 Al$_{14.9}$Zn$_{42.1}$Mg$_{43.0}$ AC. 

The narrow pseudogap in the 2/1 AC seems to be consistent with the experimental observation of the electrical resistivity, which increases with decreasing temperature from the room temperature down to around $T_c$, whereas it monotonically decreases for the 1/1 AC\cite{Kamiya}. To be more conclusive, we need an explicit calculation of the electrical conductivity using the Boltzmann equation where we may use the energy band dispersions obtained from the present calculations. In addition to the electrical conductivity, it is interesting to calculate the other transport coefficients such as the Hall coefficient and the Seebeck coefficient, using the same approach and to compare the results between the 1/1 and 2/1 ACs. 

There are about $5\%$ discrepancies between the calculated and experimental values of $\gamma$ for the both 1/1 and 2/1 ACs.  This may be due to the slight difference in the composition between the calculated and experimental systems. In addition, the effects of the chemical disorder in Zn/Al atoms are considered to occur in the real materials but are not considered in the present calculations. The virtual crystal approximation is an available approach to consider such effects on the basis of the first-principles calculation and will be an important future work.  

As the local structure of the 2/1 AC is very close to the QC, the electronic states of the 2/1 AC are expected to be a good starting point to discuss the QC. In fact, the temperature dependence of the electrical resistivity and the value of $T_c$ in the 2/1 AC almost coincide with those in the QC. Therefore, it is important to calculate the transport properties and the superconductivity in the 2/1 AC on the basis of the first-principles calculation to also discuss those in the QC. Explicit calculations are now under way and will be reported elsewhere.

\section{Acknowledgment}
This work was partially supported by a Grant-in-Aid for Scientific Research from the Ministry of Education, Culture, Sports, Science and Technology. Numerical calculations were performed in part using OFP at the CCS, University of Tsukuba, MASAMUNE-IMR at the CCMS, Tohoku University, and the Supercomputer Center at the ISSP, the University of Tokyo.

\end{document}